\documentclass[10pt, conference, letterpaper]{IEEEtran}
\pdfoutput=1
\usepackage[T1]{fontenc}
\usepackage{cite}
\usepackage[pdftex]{graphicx}
\usepackage{array}
\usepackage{amsmath,amssymb,latexsym,amsmath,euscript,mathrsfs}
\usepackage{mdwtab}
\usepackage[caption=false,font=footnotesize]{subfig}
\usepackage{url}
\usepackage[ruled,vlined]{algorithm2e}

\usepackage{comment}
\usepackage{graphicx,epstopdf}

\usepackage{amsthm}
\newtheorem{theorem}{Theorem}

\newtheorem{lemma}[theorem]{Lemma}

\begin{document}
%
\title{Online VNF Scaling in Datacenters}
\author{\IEEEauthorblockN{Xiaoke Wang\IEEEauthorrefmark{1},
Chuan Wu\IEEEauthorrefmark{1},
Franck Le\IEEEauthorrefmark{2},
Alex Liu\IEEEauthorrefmark{3},
Zongpeng Li\IEEEauthorrefmark{4} and
Francis Lau\IEEEauthorrefmark{1}}
\IEEEauthorblockA{\IEEEauthorrefmark{1}The University of Hong Kong, Email: \{xkwang, cwu, fcmlau\}@cs.hku.hk}
\IEEEauthorblockA{\IEEEauthorrefmark{2}IBM T. J. Watson, Email: fle@us.ibm.com}
\IEEEauthorblockA{\IEEEauthorrefmark{3}Michigan State University, Email: alexliu@cse.msu.edu}
\IEEEauthorblockA{\IEEEauthorrefmark{4}University of Calgary, Email: zongpeng@ucalgary.ca}}

\maketitle

\begin{abstract}
Network Function Virtualization (NFV) is a promising technology that promises to significantly reduce the operational costs of network services by deploying virtualized network functions (VNFs) to commodity servers in place of dedicated hardware middleboxes. The VNFs are typically running on virtual machine instances in a cloud infrastructure, where the virtualization technology enables dynamic provisioning of VNF instances, to process the fluctuating traffic that needs to go through the network functions in a network service. In this paper, we target dynamic provisioning of enterprise network services - expressed as one or multiple service chains - in cloud datacenters, and design efficient online algorithms without requiring any information on future traffic rates. The key is to decide the number of instances of each VNF type to provision at each time, taking into consideration the server resource capacities and traffic rates between adjacent VNFs in a service chain. In the case of a single service chain, we discover an elegant structure of the problem and design an efficient randomized algorithm achieving a e/(e-1) competitive ratio. For multiple concurrent service chains, an online heuristic algorithm is proposed, which is O(1)-competitive. We demonstrate the effectiveness of our algorithms using solid theoretical analysis and trace-driven simulations.
\end{abstract}

\IEEEpeerreviewmaketitle

\section{Introduction}
\label{sec:introduction}


The newly emerged paradigm of Network Function Virtualization (NFV) aims to consolidate network functions onto industry-standard high-volume servers, switches and storage using standard IT virtualization technology, 
  in order to enable rapid network service composition/innovation, energy reduction and cost minimization for network operators \cite{isg2013white}. A network function deployed in the NFV environment is termed virtualized network function (VNF) \cite{virtualisation2014terminology}.

Virtualization of enterprise network functions has been among the initially proposed important use cases of NFV. Many enterprises require a significant number of network functions (comparable to the number of access routers), which typically constitute one or multiple network function service chains, to support their services \cite{sekar2012design}. For example, a common service chain for access service may contain an intrusion detection device, a firewall, and a load balancer that distributes incoming traffic to a pool of servers. 

A common NFV proposal is to implement an enterprise's network service chains using VNF instances running on virtual machines (VMs) in a cloud computing platform, {\em i.e.}, the enterprise's datacenter or a public cloud, where the VNFs are made available to the enterprise in the fashion of Software as a Service \cite{etsi2013network}. Such virtualization enables dynamic provisioning of VNF instances with enough resources (CPU, memory, bandwidth, etc.) to serve varying input traffic while preserving the provisioning cost. No matter whether the datacenter is owned by the enterprise or operated by a 3rd-party cloud provider, cost minimization is among the top priorities. The main provisioning cost of VNF instances is due to power consumption to operate servers and cooling facilities, which is largely decided by the numbers and the types of VMs running different VNFs. On the other hand, launching new VNF instances on servers involves transferral of VM images, booting and attaching them to devices. This leads to a deployment cost, which is typically considered on the order of the cost to run a server for a number of seconds or minutes \cite{lin2013dynamic}. Such a deployment cost should be minimized as well, by avoiding frequent deployment and removal of VNF instances.



This paper addresses the following key problem in dynamic VNF provisioning in a cloud datacenter:

{\em Given one or multiple service chains that constitute an enterprise's network service(s), how can we design online solutions that dynamically scale VNF instances and pack them onto servers, to adequately serve fluctuating input traffic, and to achieve close-to-offline-minimum provisioning cost over the long run of the system?}

The key technical challenge in designing an efficient online algorithm to fulfil the above goal lies in the unknown nature of future traffic rates arriving at each service chain, which can vary significantly over time. We have to strategically make the deployment decisions on the go, avoiding undesirable situations of launching a VNF instance immediately following destroying the previous one.

Our contributions in efficient online VNF provisioning algorithm design are summarized as follows:

{\em First}, in the case of a single service chain, we 
 present a randomized online algorithm which returns a solution of total cost at most $e/(e-1)$ times the offline optimal solution. The algorithm proceeds in two stages. The first stage is called ``pre-planning'' which prevents VNF migration across servers over time. The second stage is an adaptation of the classic ski-rental algorithm which leads to an optimal competitive ratio.

{\em Next}, in the case of multiple concurrent service chains, we show the algorithm used in the single service chain cannot be directly applied to this scenario. Instead, we present a heuristic algorithm which relies on a minimal weight matching algorithm to minimize the deployment cost and achieves a competitive ratio of O(1).

{\em Furthermore}, our algorithms are simple and easy to implement in practice in an online manner. We demonstrate the effectiveness of our algorithms using both solid theoretical analysis as well as extensive trace-driven evaluation.

\section{Related Work}\label{sec:relatedwork}

\subsection{Enabling Technologies for NFV}

There have been recent work on building efficient software middleboxes approaching performance of hardware middleboxes, {\em e.g.}, ClickOS \cite{martins2014clickos} and CoMb \cite{sekar2012design}. ClickOS is able to saturate a 10Gb pipe on a commodity server and CoMb can reduce the network provisioning cost by 60\%.

Some other work focus on management of VNF instances. SIMPLE \cite{qazi2013simple} implements an SDN-based policy enforcement layer for efficient middlebox-specific traffic steering. APLOMB \cite{sherry2012making} facilitates outsourcing of enterprise middleboxes to the cloud. Clayman {\em et al.}~\cite{clayman2014dynamic} design an orchestrator-based architecture for automatic placement of virtual nodes and allocation of network services. These systems greatly facilitate the deployment of VNF instances.

To support VNF scaling, Split/Merge \cite{rajagopalan2013split} enables efficient, load-balanced elasticity for scaling virtual middleboxes. 
 OpenNF \cite{gember2014opennf} supports loss-free and order-preserving flow state migration by leveraging events triggered by VNFs, buffering the relevant packets, and enabling a two-phase forwarding state update. We assume exploiting such a system for implementing scaling instructed by our online algorithms.

\subsection{Optimal Placement of VNFs}
There have been a few recent studies on optimizing orchestration and placement of VNFs. Stratos \cite{gember2013stratos} orchestrates the VNFs to optimize resource usage in three steps. VNF-P \cite{moens2014vnf} presents an optimization model for VNF resource allocation and designs a fast heuristic algorithm. Bari {\em et al.}~\cite{2015arXiv150306377F} study a similar problem, formulate the problem into an integer linear program (ILP) and solve it by a standard solver. Mehraghdam {\em et al.}~\cite{mehraghdam2014specifying} propose a context-free language to describe the service chains and model a mixed integer quadratic constrained program (MIQCP) to pursue different optimization goals in VNF placement. Cohen {\em et al.}~\cite{place14vnf} investigate a NFV placement problem to optimize the distance cost between clients and the virtual functions that they need. Except \cite{place14vnf}, the other studies propose heuristic algorithms to solve the respective VNF placement problems, without providing any theoretical performance guarantee. In addition, all of these work focus on offline/one-time VNF placement and resource allocation, instead of dynamic VNF provisioning, which calls for efficient online algorithm design. 



\section{Problem Model}\label{sec:formula}

\subsection{System Model}
We consider an enterprise deploying $S$ service chains in a cloud datacenter. Each service chain $s$ is an ordered sequence of VNFs. The set of links interconnecting VNFs in $s$ is denoted by $\mathbb{L}^{s}$, where $(i,j) \in \mathbb{L}^{s}$ if VNF $i$ is the predecessor of VNF $j$ in service chain $s$. For ease of problem formulation, special VNF types named VNF $0$ and $0'$ are added to the head and the tail of a service chain to indicate source and destination of the traffic flow, respectively, and the corresponding links $(0,i)$ and $(j,0')$ are added to $\mathbb{L}^{s}$ as well, assuming $i$ and $j$ are the first and the last VNF in $s$. 
 The system works in a time slotted fashion, spanning time slots $1, 2, \ldots, T$. The input traffic rate to each service chain $s$ varies from time to time, denoted by $\alpha^{s}(t)$ ({\em e.g.}, in Mbps) at time $t$.


There are in total $I$ different types of VNFs in the system. Multiple VM instances can be provisioned in the cloud datacenter to run the same type of VNF, which we refer to as \emph{instances of the VNF}. These VMs have specific configurations and require fixed resources. An instance of VNF $i$ consumes $c_{ir}$ of resource $r\in R$, where $R$ is the number of resource types including CPU, memory, storage and network bandwidth. Based on its resource composition, an instance of VNF $i$ can maximally process traffic at the rate of $b_i$ (in Mbps) at each time, without incurring prolonged packet queueing delays that violate a preset performance threshold.



There are $U$ servers in the cloud datacenter. Without loss of generality, we assume that the servers are homogeneous, each having a capacity $C_r$ of resource $r\in R$. 



\subsection{Problem Formulation}

Our dynamic VNF provisioning problem is to find the optimal numbers and server placement of VNF instances for the service chains in each time slot, in order to minimize the provisioning cost over the entire system span $T$. 
 We define the following decision variables: (i) VNF placement variable $x_{ui}(t)$, which denotes the number of instances of VNF $i$ running on server $u$ in $t$, $\forall u\in U, i\in I, t\in T$; (ii) routing variable $y_{uivj}^s(t)$, which represents the amount of traffic (in Mbps) in service chain $s$, forwarded from instances of VNF $i$ on server $u$ to instances of VNF $j$ on server $v$, $\forall u,v\in U, i,j\in I, s\in S, t\in T$. 

The VNF provisioning cost contains two parts:

{\bf I. VNF operational cost.} Let $\phi_i$ denote the cost of running an instance of VNF $i$ per time slot, mainly attributed to the power consumption of the hosting server. The overall operational cost is 
\begin{equation}\label{eq:VNF service cost}
\mathbb{O} = \sum\limits_{t\in [T]}\sum\limits_{u\in [U]}\sum\limits_{i\in [I]}\phi_ix_{ui}(t)
\end{equation}

{\bf II. VNF deployment cost.} 
 Deploying a new instance of VNF $i$ requires transferring a VM image containing the network function, booting it and attaching it to devices on the server. We associate a deployment cost $\varphi_i$ with the process. The overall VNF deployment cost 
can be expressed as
\begin{equation}\label{eq:VNF deployment cost}
\mathbb{D} = \sum\limits_{t\in [T]}\sum\limits_{u\in [U]}\sum\limits_{i\in [I]}\varphi_i[x_{ui}(t)-x_{ui}({t-1})]^+
\end{equation}

\begin{align*}
\text{where}[x_{ui}(t)-x_{ui}({t-1})]^+ & = \max{\{x_{ui}(t)-x_{ui}({t-1}), 0}\}
\end{align*}

indicates the number of newly added instances of VNF $i$ on server $u$ in $t$.

Our objective is to minimize the total cost $\mathbb{O}+\mathbb{D}$. The constraints that the decision variables should respect include the following.

{\em First}, all incoming traffic to each service chain at each time should be served. This can be guaranteed by the following constraints.

(i) Taking in all the incoming traffic to instances of the first VNF in a service chain (possibly deployed in different servers):
\begin{equation}\label{eq:input}
\sum\limits_{v\in [U]}\sum_{j:(0,j)\in \mathbb{L}^s}y_{00vj}^s(t) \geq \alpha^s(t), \forall s \in [S], t \in [T]
\end{equation}
\noindent where $y_{00vj}^s(t)$ denotes the incoming traffic rate (from the dummy VNF $0$ on an imaginary server $0$) directed to instances of VNF $j$ on server $v$, which is summed up in the left-hand side of (\ref{eq:input}) only if VNF $j$ is the first VNF in the chain, {\em i.e.}, $(0,j)\in \mathbb{L}^s$.

(ii) Flow conservation at instances of each VNF $i$ in service chain $s$ on each server $u$:
\begin{equation}\label{eq:flow conservation}
\begin{split}
\sum\limits_{v\in [U]}\sum_{(i,j)\in \mathbb{L}^s}y_{uivj}^s(t) = \lambda_i^s\sum\limits_{v\in [U]}\sum_{(j,i)\in \mathbb{L}^s}y_{vjui}^s(t), \\
\forall s \in [S], u \in [U], i \in [I], t \in [T]
\end{split}
\end{equation}
\noindent where the left-hand-side is the overall outgoing traffic from instances of VNF $i$ on server $u$ to instances of next-hop VNF $j$ of service chain $s$ on different servers, and the summation in the right-hand side computes the overall incoming traffic to instances of VNF $i$ on $u$ from instances of the previous-hop VNF deployed on different servers.
It is worth noting that the flow rate of traffic may change after being processed by a network function \cite{gember2013stratos}. 
Hence we define a gain/drop factor $\lambda_i^s$ for each VNF $i$ in each service chain $s$.

(iii) Provisioning a sufficient number of instances of each VNF on each server, whose overall processing capacity can serve all the received flows (possibly) from different previous-hop VNFs in different service chains:
\begin{equation}\label{eq:VNF processing capacity}
\sum\limits_{s\in[S]}\sum\limits_{v\in [U]}\sum_{(j,i)\in \mathbb{L}^s}y_{vjui}^s(t) \leq x_{ui}(t)b_i, \forall i \in [I], u \in [U], t \in [T]
\end{equation}

{\em In addition}, resource capacities on each server should not be over-committed by the deployed VNF instances at any time, as expressed by the following constraints:
\begin{equation}\label{eq:server CPU}
\sum\limits_{i\in [I]}x_{ui}(t)c_{ir} \leq C_r, \forall u \in [U], r \in [R], t \in [T]
\end{equation}

In summary, the offline VNF provisioning problem is formulated as follows:
\begin{equation}\label{eq1:total_cost}
\min   \sum\limits_{t\in [T]}\sum\limits_{u\in [U]}\sum\limits_{i\in[I]}(\phi_ix_{ui}(t) + \varphi_i[x_{ui}(t)-x_{ui}({t-1})]^+) \\
\end{equation}

s.t. ~~~constraints (\ref{eq:input})(\ref{eq:flow conservation})(\ref{eq:VNF processing capacity})(\ref{eq:server CPU})
\begin{equation}
x_{ui}(t) \in \{0,1,...\}, \forall u\in [U], i\in [I], t\in[T]
\end{equation}
\begin{equation}
~~~~~~~~~y_{uivj}^s(t) \geq 0, \forall s\in[S], u,v\in [U], i,j\in [I], t\in[T]
\end{equation}

We list important notation in this paper in Table \ref{tb:notation}, for ease of reference.

\begin{table}
\begin{center}
\begin{tabular}{|c|l|}
\hline
$[X]$ & integer set \{1,2,...,X\}\\
\hline
$U$ & \# of servers in the system\\
\hline
$T$ & \# of time slots\\
\hline
$I$ & \# of VNF types\\
\hline
$S$ & \# of service chains\\
\hline
$R$ & \# of resource types in a server\\
\hline
$\mathbb{L}^s$ & the set of links between VNFs in service chain $s$\\
\hline
$x_{ui}(t)$ & \# of VNF $i$ on server $u$ in time slot $t$\\
\hline
$y_{uivj}^s(t)$ & traffic from VNF $i$ on server $u$ to VNF $j$ on server $v$ in $t$\\
\hline
$\alpha^s(t)$ & the input traffic to service chain $s$ at time $t$\\
\hline
$\lambda_i^s$ & gain/drop factor of VNF $i$ in service chain $s$\\
\hline
$c_{ir}$ & VNF $i$'s consumption of resource $r$\\
\hline
$C_{r}$ & capacity of resource $r$ on a server\\
\hline
$b_{i}$ & processing capacity of an instance VNF $i$\\
\hline
$\phi_i$ & per-time-slot operational cost of an instance of VNF $i$\\
\hline
$\varphi_i$ & deployment cost of an instance of VNF $i$\\
\hline
\end{tabular}
\end{center}
\caption{Notation}
\label{tb:notation}
\end{table}


\subsection{Simplifying the Offline Optimization Problem}

The offline optimization problem in (\ref{eq1:total_cost}) gives the optimal VNF deployment and traffic routing decisions at each time, using complete information of traffic rates during the entire system span. Towards designing an efficient online algorithm, we study the structure of the offline problem carefully, and identify that the problem formulation can actually be simplified by removing routing decisions $y_{uivj}^s(t)$'s, leaving only VNF placement decisions $x_{ui}(t)$'s. The key observation is that given specific input traffic rates to the service chains at $t$, $\alpha^{s}(t),\forall s\in[S]$, the minimal number of instances of each VNF required in the entire system at the time can be obtained based on constraints (\ref{eq:input}), (\ref{eq:flow conservation}) and (\ref{eq:VNF processing capacity}), as stated in the following theorem.

\begin{theorem} \label{thm:lower bounds of VNFs}
The minimal number of instances of VNF $i$ required to serve input traffic rates $\alpha^s(t),\forall s\in[S]$, in time slot $t$ is $n_i(t) = \lceil \frac{\sum_{s\in[S]}\bar{\lambda}_i^s\alpha^s(t)}{b_i} \rceil$, where $\bar{\lambda}_i^s=\prod\limits_{j=1}^{N^s(j+1)=i}\lambda_{N^s(j)}^s$ is the cumulative gain/drop factor in service chain $s$ before the flow enters VNF $i$ in the chain and $N^s(j)$ denotes the type of j-th VNF in the service chain $s$.
\end{theorem}

The detailed proof is given in Appendix \ref{app:thm:lower_bounds_of_VNFs}.

Then we can convert the offline problem (\ref{eq1:total_cost}) into the following one:

\begin{equation}\label{eq2:total_cost}
\min   \sum\limits_{t\in [T]}\sum\limits_{u\in [U]}\sum\limits_{i\in[I]}(\phi_ix_{ui}(t) + \varphi_i[x_{ui}(t)-x_{ui}({t-1})]^+) \\
\end{equation}

s.t. ~~~~~ \begin{equation}\label{eq2:input}
\sum\limits_{u\in [U]}x_{ui}(t) \geq n_i(t), \forall i \in [I], t \in [T]
\end{equation}
\begin{equation}\label{eq2:server CPU}
\sum\limits_{i\in [I]}x_{ui}(t)c_{ir} \leq C_r, \forall u \in [U], r \in [R], t \in [T]
\end{equation}
\begin{equation}
x_{ui}(t) \in \{0,1,...\}, \forall u\in [U], i\in [I], t\in[T]
\end{equation}

In fact, we are able to prove the following theorem:

\begin{theorem} \label{thm:equal_offline_problems}
The offline VNF provisioning problem in (\ref{eq2:total_cost}) is equivalent to the offline problem in (\ref{eq1:total_cost}).
\end{theorem}
The detailed proof is given in Appendix \ref{app:thm:equal_offline_problems}.

In what follows, we design online algorithms to make VNF placement and traffic routing decisions on the go, based only on current information and past history. We divide our design into two cases, for deployment of a single service chain and concurrent deployment of multiple service chains, respectively.

\section{Online Algorithm for a Single Service Chain}\label{sec:SSC}

In this section, we focus on a single service chain containing VNFs $1\rightarrow 2\rightarrow \cdots \rightarrow L$. We remove $s$ (that indicates a service chain) from all relevant notation, for simplicity.

\subsection{Insights from the Offline Problem}
\label{sec:ssc_insights}
The offline VNF provisioning problem in a system with only one service chain takes the same form as problem (\ref{eq2:total_cost}), except that the minimal number of instances of each VNF $i$ in constraint (\ref{eq2:input}) is computed as $n_i(t) = \lceil \frac{\bar{\lambda}_i\alpha(t)}{b_i} \rceil$. 

We observe that the objective function is equivalent to $\sum\limits_{t\in [T]}\sum\limits_{i\in[I]}(\sum\limits_{u\in [U]}\phi_ix_{ui}(t) + \sum\limits_{u\in [U]}\varphi_i[x_{ui}(t)-x_{ui}({t-1})]^+)$. In the second term, we have $\sum\limits_{u\in [U]}[x_{ui}(t)-x_{ui}({t-1})]^+ \geq [\sum\limits_{u\in [U]}x_{ui}(t)-\sum\limits_{u\in [U]}x_{ui}({t-1})]^+ $. The equality holds if and only if the signs of $x_{ui}(t)-x_{ui}({t-1}),\forall u\in[U]$, are all the same: from $t-1$ to $t$, the numbers of instances of VNF $i$ deployed on all the servers either all increase ($x_{ui}(t)\ge x_{ui}({t-1}),\forall u\in[U]$) or all decrease ($x_{ui}(t)\le x_{ui}({t-1}),\forall u\in[U]$); there does not exist a pair of servers $u_1$ and $u_2$, such that $x_{u_1i}(t)> x_{u_1i}({t-1})$ and $x_{u_2i}(t)< x_{u_2i}({t-1})$. The later can be considered as a {\em VNF instance migration} case, {\em i.e.}, at least one instance of VNF $i$ is moved from one server to another.


The following is an example where VNF instance migration has to be done due to limited capacity. Suppose there are 2 servers, each of which has 1 unit resource. The service chain is VNF $1\rightarrow 2$. An instance of VNF 1 requires 0.5 unit of resource and an instance of VNF 2 requires 0.2 unit. At first, there is 1 instance of VNF 1 and 1 instance of VNF 2 on each server. Then, due to the increase of input traffic rate, we have to add 1 instance for each VNF. However, none of the two servers has enough resource to accommodate an additional instance of VNF 1. We have to migrate one instance of VNF 2 from one server to the other to make room for the additional instance of VNF 1.


Nevertheless, if we can somehow ensure that no migration is needed (we will design an algorithm for it), the objective function is equivalent to $\sum\limits_{t\in [T]}\sum\limits_{i\in[I]}(\sum\limits_{u\in [U]}\phi_ix_{ui}(t) + \varphi_i[\sum\limits_{u\in [U]}x_{ui}(t)-\sum\limits_{u\in [U]}x_{ui}({t-1})]^+)$. Furthermore, if the server capacity constraint (\ref{eq2:server CPU}) can be ignored temporarily, then we can create a new decision variable $x_i(t) = \sum\limits_{u\in[U]}x_{ui}(t)$, denoting the total number of instances of VNF $i$ deployed in the system in $t$, and convert the offline problem (\ref{eq2:total_cost}) into the following:

\begin{equation}\label{eq3:total_cost}
\min   \sum\limits_{t\in[T]}\sum\limits_{i\in[I]}(\phi_ix_{i}(t) + \varphi_i[x_{i}(t)-x_{i}({t-1})]^+) \\
\end{equation}

s.t.
\begin{equation}\label{eq3:input}
x_{i}(t) \geq n_i(t), \forall i \in [I], t \in [T]
\end{equation}
\begin{equation}
x_{i}(t) \in \{0,1,...\}, \forall i\in[I], t \in [T]
\end{equation}

The optimization problem (\ref{eq3:total_cost}) deals with the total number of instances to deploy for each VNF type to fulfil the traffic demand at each time, without taking care of the detailed placement of the VNF instances on servers, {\em i.e.}, how many of $x_{i}(t)$ to place on a server $u$. In addition, there is no coupling of decisions among different types of VNFs, such that we can separately optimize the total number of instances to deploy for each VNF, by solving the subproblem:

\begin{equation}\label{eq4:total_cost}
	\min   \sum\limits_{t\in[T]}(\phi_ix_{i}(t) + \varphi_i[x_{i}(t)-x_{i}({t-1})]^+)\\
\end{equation}

s.t.
\begin{equation}
	x_{i}(t) \geq n_i(t), x_{i}(t) \in \{0,1,...\}, \forall t \in [T]\nonumber\\
\end{equation}	
	
\noindent We observe that the offline optimization problem (\ref{eq4:total_cost}) can be solved by doing some minor modifications to classic ski-rental algorithms \cite{karlin1994competitive}. It can be verified that if in the next $\lfloor\varphi_i/\phi_i\rfloor$ time slots, the number of VNF $i$ instances required is smaller or equal to current number of VNF $i$ instances, then the most economical way is to remove the extra VNF $i$ instances and only keep the maximum number of VNF $i$ instances required by the next $\lfloor\varphi_i/\phi_i\rfloor$ time slots. This is exactly the idea behind the offline ski-rental algorithms. Therefore, to design an online algorithm for problem (\ref{eq4:total_cost}), we can apply the classic ski-rental algorithms, such as that in \cite{karlin1994competitive}. We distinguish idle VNF instances from running VNF instances. For idle instances, they would not be removed from servers until the accumulated operational cost is larger than their deployment cost. If an idle instance is required during that time duration, we simply turn its state into running without causing any deployment cost. The detailed algorithm is to be discussed in Alg.~\ref{alg:online randomized}.

The pending issue for designing an online algorithm that solves (\ref{eq2:total_cost}), is to come up with an efficient VNF placement scheme, which places VNF instances at the total number of $x_{i}(t)$ from (\ref{eq3:total_cost}) on the servers, ensuring that server capacity constraint in (\ref{eq2:server CPU}) is respected, and no VNF instance migration would happen, {\em i.e.}, from $t-1$ to $t$, the servers are all hosting more instances of VNF $i$ or are all hosting less instances of VNF $i$. 
In the next subsection, we propose an efficient pre-planning step, which produces a feasible VNF placement solution, assuming that the input traffic rate to the service chain is the largest possible that the system can support. We then design our online algorithm to solve (\ref{eq2:total_cost}), combining $x_{i}(t)$'s computed from (\ref{eq3:total_cost}) and the VNF placement scheme derived from the pre-planning step, which satisfies the above conditions.

\subsection{VNF Deployment for Maximum Traffic Rate}

We now investigate the maximum input traffic rate $\alpha^{max}$ that the system can support in a single time slot, as well as obtain a feasible VNF deployment solution to serve a flow at this maximum rate. 
 Given an input traffic rate $\alpha$, we can obtain the corresponding minimum number of instances of each VNF needed, $n_i$, according to Theorem \ref{thm:lower bounds of VNFs}, for all $i$ in the service chain. Then whether $U$ servers are enough to host these VNFs can be determined by finding a feasibility solution following optimization problem, where binary variable $z_{u}$ denotes whether server $u$ is used or not:


%
\begin{equation}\label{eq01:input}
\sum\limits_{u\in [U]}x_{ui} \geq n_{i}, \forall i \in [L]
\end{equation}
\begin{equation}\label{eq01:server CPU}
\sum\limits_{i\in [l]}x_{ui}c_{ir} \leq C_rz_{u}, \forall u \in [U], r \in [R]
\end{equation}
\begin{equation}
x_{ui} \in \{0,1,...\}, z_{ui} \in \{0,1\}, \forall u\in[U], i\in[L]
\end{equation}


To decide the maximum supportable input traffic rate, we can apply a bi-section algorithm. The complete algorithm is given in Alg.~\ref{alg:MITR}.



\begin{algorithm}	
	\caption{Pre-planning Step}\label{alg:MITR}
    \small
	
        \KwIn{$U, L, R, \mathbf{\lambda}, \mathbf{b}, \mathbf{c}, \mathbf{C}$}
        \KwOut{$\alpha^{max}$, $\mathbf{x}^{max}$}
        \BlankLine
        $l_b = 0, u_b = MAXRATE$;\\
        \While{$u_b - l_b > 1$}{
        $m := (u_b + l_b) / 2$;\\
		Solve problem (\ref{eq01:input}) by setting $n_i= \lceil \frac{\bar{\lambda}_im}{b_i} \rceil$, using bin-packing algorithm in \cite{goemans2014polynomiality}\\
        \eIf    {Exists a feasible solution $\mathbf{x}$ to problem (\ref{eq01:input})}{
        $l_b := m$
        }{
        $u_b := m$
        }
        }
        $\alpha^{max} := l_b$, $\mathbf{x}^{max} := \mathbf{x}$
		
\end{algorithm}

In Alg.~\ref{alg:MITR}, we set the initial search range for $\alpha^{max}$ to $[0, MAXRATE]$, where $MAXRATE$ can be a very large estimated rate.
To solve problem (\ref{eq01:input}) at specific $n_i$'s, we note that the problem is the high multiplicity variant of a multi-dimensional bin packing problem \cite{coffman1996approximation}, which is NP hard if the length $L$ of the service chain is a large number. However, the length of a service chain is usually short - the length of most of the representative NFV service chains is no larger than $4$. Hence we take $L$ as a fixed small constant. In this case, we show that there exists a polynomial-time algorithm to solve the problem. In particular, Goemans and Rothvoss \cite{goemans2014polynomiality} show that the 1-dimensional bin packing problem with a fixed number of item types ($L$ in our model) and high multiplicity can be solved in polynomial time. We prove that the same algorithm from \cite{goemans2014polynomiality} can be applied to solve bin packing problems of any fixed dimensions, {\em i.e.}, the number of resource types $R$ in our model, in polynomial time as Theorem \ref{thm:polynomial} shows.



\begin{theorem} \label{thm:polynomial}
Alg.~\ref{alg:MITR} runs in polynomial time. 
\end{theorem}
The detailed proof is given in Appendix \ref{app:thm:polynomial}.

\subsection{VNF Placement Scheme}
 Combining the pre-planning step in Alg.~\ref{alg:MITR} with our insights in Sec.~\ref{sec:ssc_insights}, we are able to design an efficient VNF placement scheme which can guarantee server capacity constraint (\ref{eq2:server CPU}) is respected and no migration of VNF instances occurs. The main idea is to guarantee the set of server placement of VNF instances to serve the input traffic rate $\alpha(t)$ for any $t$ is always a subset of the set of server placement of VNF instances to serve the maximum input traffic rate $\alpha^{max}$.\footnote{ We reasonably assume the input traffic rate $\alpha(t)$ in $[0,T]$ is within $[0,\alpha^{max}]$, {\em i.e.}, the datacenter's capacity is sufficient to serve the input traffic.} For example, suppose we have 3 instances of VNF $1$  on server 1 and 1 instance of VNF $1$ on server 2 to serve $\alpha^{max}$; then to serve $\alpha(t)$, we could deploy 2 instances of VNF $1$ on server 1 but never deploy 2 instances of VNF $1$ on server 2. Such a placement scheme always respects server capacity constraint (\ref{eq2:server CPU}) since the VNF placement to serve $\alpha^{max}$ is governed by the constraint already.




To accomplish the above, all we need is a multiset, $S_i$ for VNF $i$, which includes the IDs of servers where all the instances of VNF $i$ are placed according to the solution to serve the maximum traffic rate. For example, if we need 2 instances of VNF $i$ on server $1$, 3 instances of VNF $i$ on server $2$ and 1 instance of VNF $i$ on server $3$ to serve the maximum rate, then $S_i = \{1,1,2,2,2,3\}$. Using $S_i$, when we need to deploy more VNF $i$ instances, we simply eject sufficient server IDs from $S_i$ and deploy VNF $i$ instances on the respective servers accordingly; when some instances of VNF $i$ are removed from the servers, we insert the respective server IDs into $S_i$. 

\begin{theorem} \label{thm:MITR}
This VNF placement scheme can guarantee server capacity constraint (\ref{eq2:server CPU}) is always respected and no migration of VNF instances occurs. 
\end{theorem}

The detailed proof is given in Appendix \ref{app:thm:MITR}.

\subsection{Online Algorithm}

Our proposed online algorithm to solve problem (\ref{eq2:total_cost}) combines the classic ski-rental algorithm from \cite{karlin1994competitive} with our proposed VNF placement scheme. More specifically, the VNF instances on the servers are either marked as running or idle. The total number of running VNF $i$ instances in time slot $t$ is $n_i(t)$ and the total number of all VNF $i$ instances on the servers in time slot $t$ is $x_i(t)$. In each time slot, if $n_i(t) \geq x_i(t-1)$, then switch all the idle VNF $i$ instances to running, eject $n_i(t)-x_i(t-1)$ elements from $S_i$ and deploy $n_i(t)-x_i(t-1)$ VNF $i$ instances on the respective servers; if $n_i(t-1)\leq n_i(t)<x_i(t-1)$, switch $n_i(t)-n_i(t-1)$ VNF $i$ instances, which become idle most recently, into running state; otherwise, switch $n_i(t-1)-n_i(t)$ running VNF $i$ instances to idle. For each idle VNF $i$ instance, we generate a ``deadline'' following a given distribution: P\{``deadline'' $= j$\} = $(\frac{\Delta_i-1}{\Delta_i})^{\Delta_i-j}\frac{1}{\Delta_i(1-(1-1/\Delta_i)^{\Delta_i})}$, where $\Delta_i = \lfloor \varphi_i/\phi_i \rfloor$. Remove the idle instance from its server completely after it has been idle for the ``deadline''number of  time slots and insert the corresponding server ID into $S_i$.

It can be proved that our Alg.~\ref{alg:online randomized} runs very fast.

\begin{theorem} \label{thm:computational complexity}
The randomized online algorithm in Alg.~\ref{alg:online randomized} has a worst-case computation complexity of $O(K)$ per time slot, where $K$ is the maximum number of VNF instances the datacenter can serve.
\end{theorem}

The competitive ratio of our Alg.~\ref{alg:online randomized} to solve problem (\ref{eq2:total_cost}) can be shown to be equal to the competitive ratio achieved by the randomized ski-rental algorithm to solve problem (\ref{eq4:total_cost}). 
\begin{theorem} \label{thm:online randomized}
The randomized online algorithm in Alg.~\ref{alg:online randomized} produces a feasible solution of (\ref{eq2:total_cost}) and achieves a competitive ratio of $e/(e-1)$.
\end{theorem}

The detailed proof is given in Appendix \ref{app:thm:online randomized}.


\begin{algorithm}[!t]
	
	\caption{Online Algorithm for Single Service Chain}\label{alg:online randomized}
    \small
	
        \KwIn{$\mathbf{n}(t)$, $\mathbf{n}(t-1)$, $\mathbf{S}$, $\mathbf{x}(t-1)$ }
        \KwOut{$\mathbf{x}(t)$}
        \BlankLine

        \For{$i \in [L]$}{
        \uIf{$n_i(t) \geq x_i(t-1)$}{
        Switch all the idle VNF $i$ instances to running;\\
        Eject $n_i(t)-x_i(t-1)$ elements from $S_i$;\\
        Place $n_i(t)-x_i(t-1)$ instances on respective servers;
        }\uElseIf{$n_i(t-1)\leq n_i(t)<x_i(t-1)$}{
        Switch $n_i(t)-n_i(t-1)$ idle VNF $i$ instances to running;
        }\Else{
        Switch $n_i(t-1)-n_i(t)$ running VNF $i$ instances to idle;
        }

        \ForAll{idle VNF $i$ instances}{
        \If{marked as running in the previous time slot}{
        counter := 0;\\
        deadline $:= j$ with probability $(\frac{\Delta_i-1}{\Delta_i})^{\Delta_i-j}\frac{1}{\Delta_i(1-(1-1/\Delta_i)^{\Delta_i})}$;}
        \If{counter $\geq$ deadline}{
        Remove it from the server;\\
        Insert the server ID into $S_i$;
        }

        }

        $x_{ui}(t)$ :=total \# of running and idle VNF $i$ instances on server $u$ at time $t$

        }

\end{algorithm}

\section{Online Algorithm for Multiple Service Chains}\label{sec:MSC}

We next focus on the general scenario where there are more than one service chains in the system.

The algorithm we designed for a single service chain cannot be extended to the case of multiple service chains, due mainly to the non-existence of the maximum input flow rate vector which uniquely decides a MAX VNF deployment solution for each service chain, {\em i.e.}, that to serve the maximal flow rates. When multiple chains coexist and may share the same VNF instances, we only have {\em pareto} maximal input flow rate vectors among the flows. Since the MAX VNF deployment solution no longer exists for each chain, we are not able to design VNF placement solution over time as subsets of the MAX placement solution, without incurring any VNF migration. Hence, we adopt a new approach.

Our analysis has revealed that any online algorithm which uses the minimal number of VNF instances at each time slot is $(1+\max\limits_{i\in[I]}\frac{\varphi_i}{\phi_i})$-competitive, as Theorem \ref{thm2:MSC competitive ratio} shows. That is, if the online algorithm finds a solution which makes $\sum\limits_{u\in [U]}x_{ui}(t) = n_i(t)$ in (\ref{eq2:input}) and further satisfies constraint (\ref{eq2:server CPU}), it is $(1+\max\limits_{i\in[I]}\frac{\varphi_i}{\phi_i})$-competitive.

\begin{theorem} \label{thm2:MSC competitive ratio}
Any online algorithm which uses the minimal number of VNF instances in each time slot and respects the server capacity constraint is $(1+\max\limits_{i\in [I]}\frac{\varphi_i}{\phi_i})$-competitive.
\end{theorem}
The detailed proof is given in Appendix \ref{app:thm2:MSC competitive ratio}.

One natural idea is to design an algorithm that packs this minimal number of VNF instances into servers, without violating capacities. Again, this can be done using the bin-packing algorithm in \cite{goemans2014polynomiality}. The bin-packing algorithm returns a set of \emph{patterns} and their corresponding \emph{multiplicities} \cite{goemans2014polynomiality}. A pattern is a feasible placement solution of instances of all VNFs in a server, which can be described by an $I$-dimensional vector $V$. The i-th component, $V(i)$ in $V$, denotes the number of VNF $i$ instances to be deployed on the server. A pattern can be applied to several servers and the number of times it is applied is called its multiplicity. However, the bin-packing algorithm does not tell us which actual server(s) is(are) to be used for deployment of each pattern.

We map the result patterns to individual servers, trying to minimize the deployment cost from one time slot to the next. 
For example, suppose in the previous time slot $t-1$, we have 2 instances of VNF 1 on server 1 and 2 instances of VNF 2 on server 2. In $t$, the bin-packing algorithm returns two patterns, (2,0) and (0,3), each of which is used once. Then we should map (2,0) to server 1 (i.e., run 2 instances of VNF 1 on server 1 in $t$) and (0,3) to server 2 (i.e., run 3 instances of VNF 2 on server 2 in $t$) . 

Formally, the deployment cost incurred is $\sum\limits_{i\in [I]}\varphi_i[V(i)-x_{ui}(t-1)]$, if we map a pattern $V$ to a server $u$. To minimize deployment cost through mapping, we have a minimum weight matching problem, where the weight on a link connecting a pattern to a server is the corresponding deployment cost. The minimum weight matching problem can be solved efficiently, {\em e.g.}, using the Kuhn-Munkres algorithm \cite{munkres1957algorithms}.

\begin{algorithm}[!t]
    \small
	\caption{Online Algorithm for Multiple Service Chains}\label{alg:online msc}

        \KwIn{$\mathbf{n}(t)$, $\mathbf{x}(t-1)$}
        \KwOut{$\mathbf{x}(t)$}
        \BlankLine

        $\mathbf{P}$ = patterns returned by Bin\_Packing($\mathbf{n}(t)$);\\
        $p$ = |$\mathbf{P}$|;\\
        $\mathbf{K}$ = pattern multiplicities returned by Bin-Packing($\mathbf{n}(t)$);\\
        Let $x_i$ denote the pattern in server $i$ in time $t-1$;
        $S_1 = \{x_1,x_2...,x_U\}$;\\
        $S_2 = \{\underbrace{P(1),...,P(1)}\limits_{K(1)},...,\underbrace{P(p),...,P(p)}\limits_{K(p)}\}$;\\
        $w_{ij} = [S_2(j)-S_1(i)]^+$;\\
        $\mathbf{x}(t)$ = Minimum\_Weight\_Matching($S_1$,$S_2$,$W$);
	
\end{algorithm}

Our online algorithm for VNF deployment of multiple service chains is given in Alg.~\ref{alg:online msc}. Note that the mapping component of the algorithm in general reduces the overall cost in (\ref{eq2:input}) achieved by the algorithm, as compared to one that does no mapping but only packs $n_i(t)$ instances of VNF $i$ into servers. 
Therefore, the competitive ratio that Alg.~\ref{alg:online msc} achieves is at most $(1+\max\limits_{i\in[I]}\frac{\varphi_i}{\phi_i})$ based on Theorem \ref{thm2:MSC competitive ratio}.

\begin{theorem}
The online algorithm in Alg.~\ref{alg:online msc} is $(1+\max\limits_{i\in[I]}\frac{\varphi_i}{\phi_i})$-competitive.
\end{theorem}


\captionsetup[figure]{labelfont=bf}
\begin{figure*}[t]
\captionsetup{width=0.48\textwidth}
\begin{center}
\begin{minipage}[t]{0.22\linewidth}

  \includegraphics[width=\textwidth]{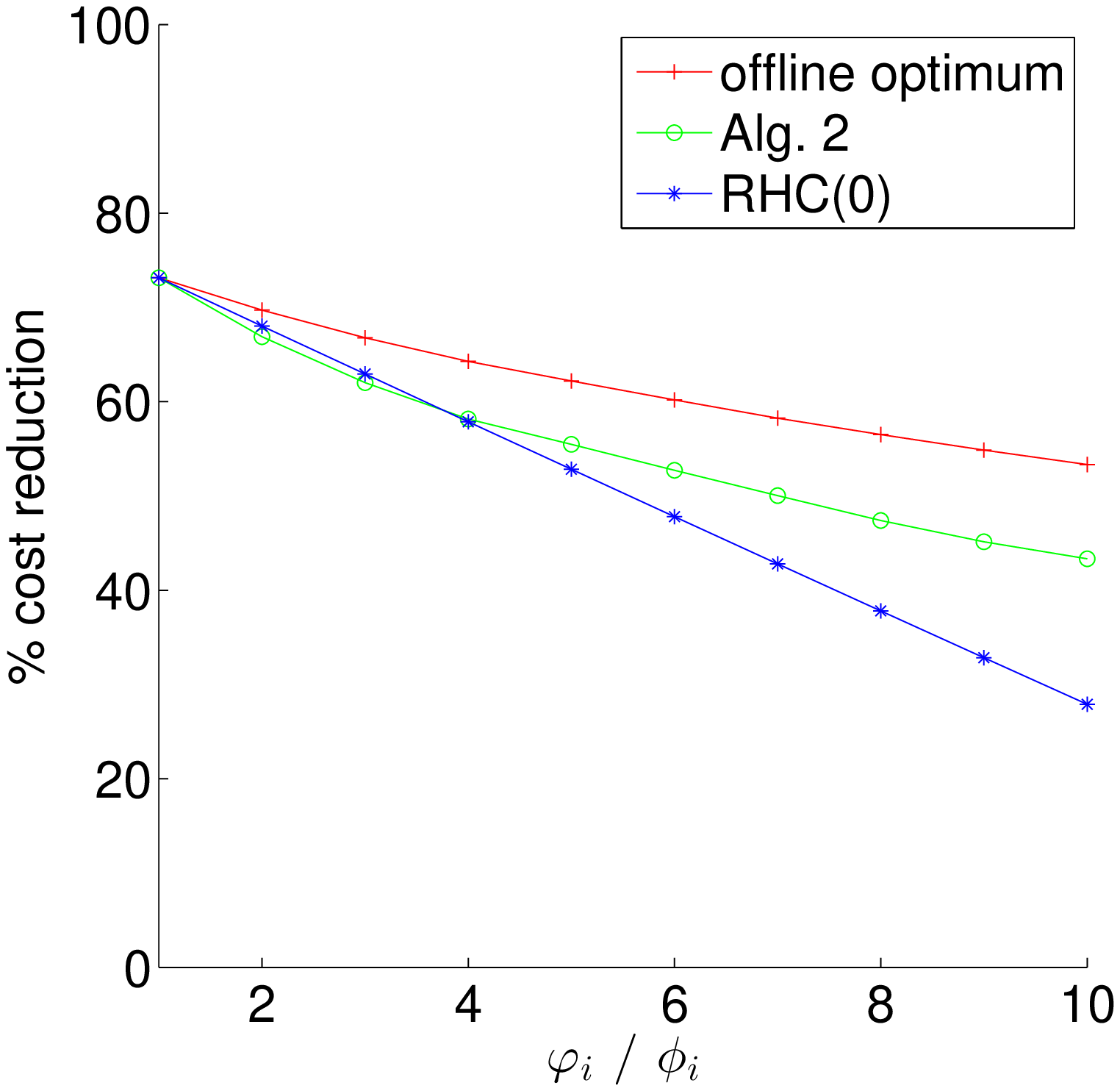}
  \caption{Impact of $\varphi_i/\phi_i$ on cost reduction for SSC}
  \label{fig:deployment2operational}
\end{minipage}
\hfill
\begin{minipage}[t]{0.22\linewidth}
  \includegraphics[width=\textwidth]{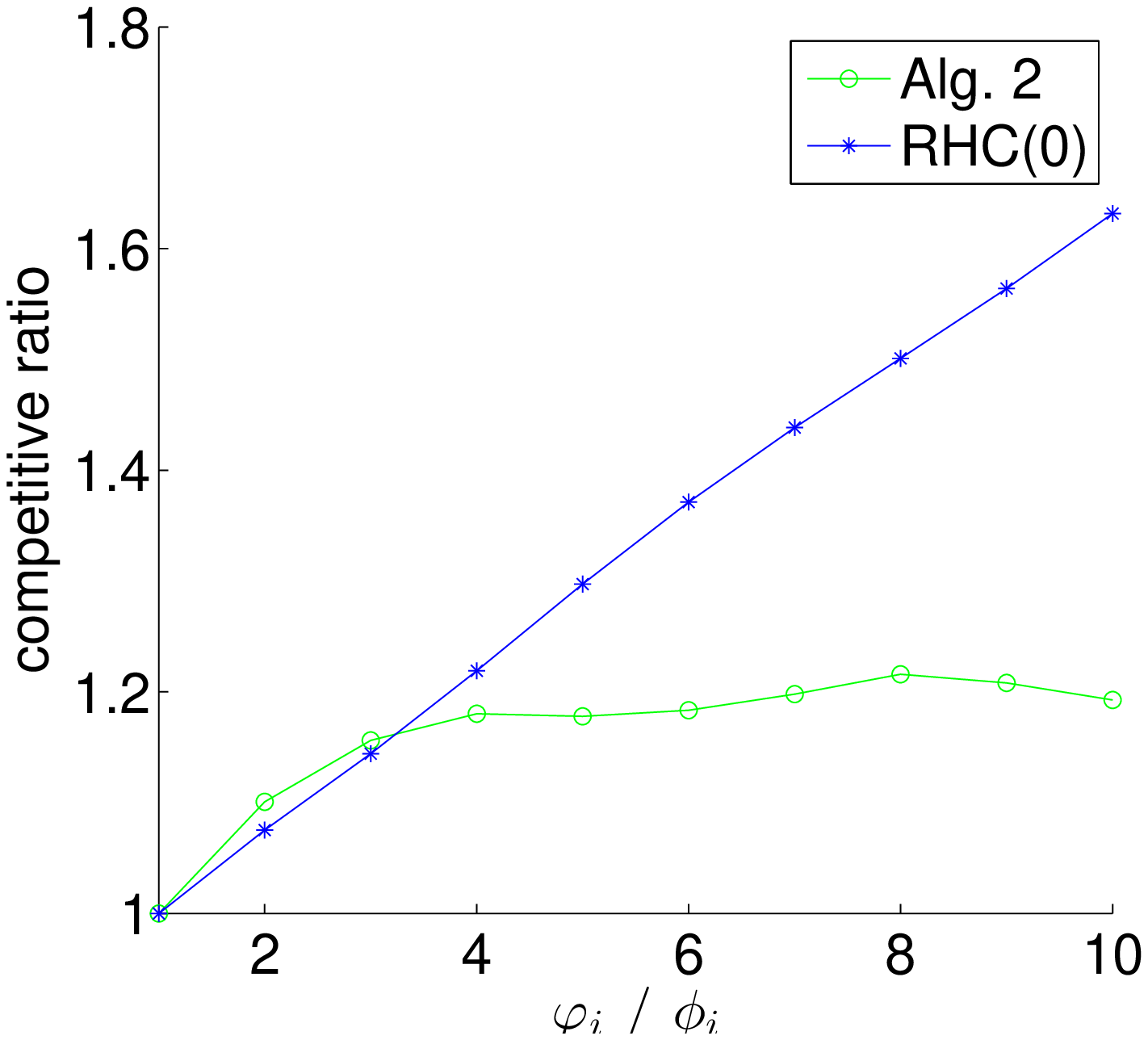}
  \caption{Impact of $\varphi_i/\phi_i$ on competitiveness for SSC}
  \label{fig:d2o_comp}
\end{minipage}
\hfill
\begin{minipage}[t]{0.22\linewidth}
  \includegraphics[width=\textwidth]{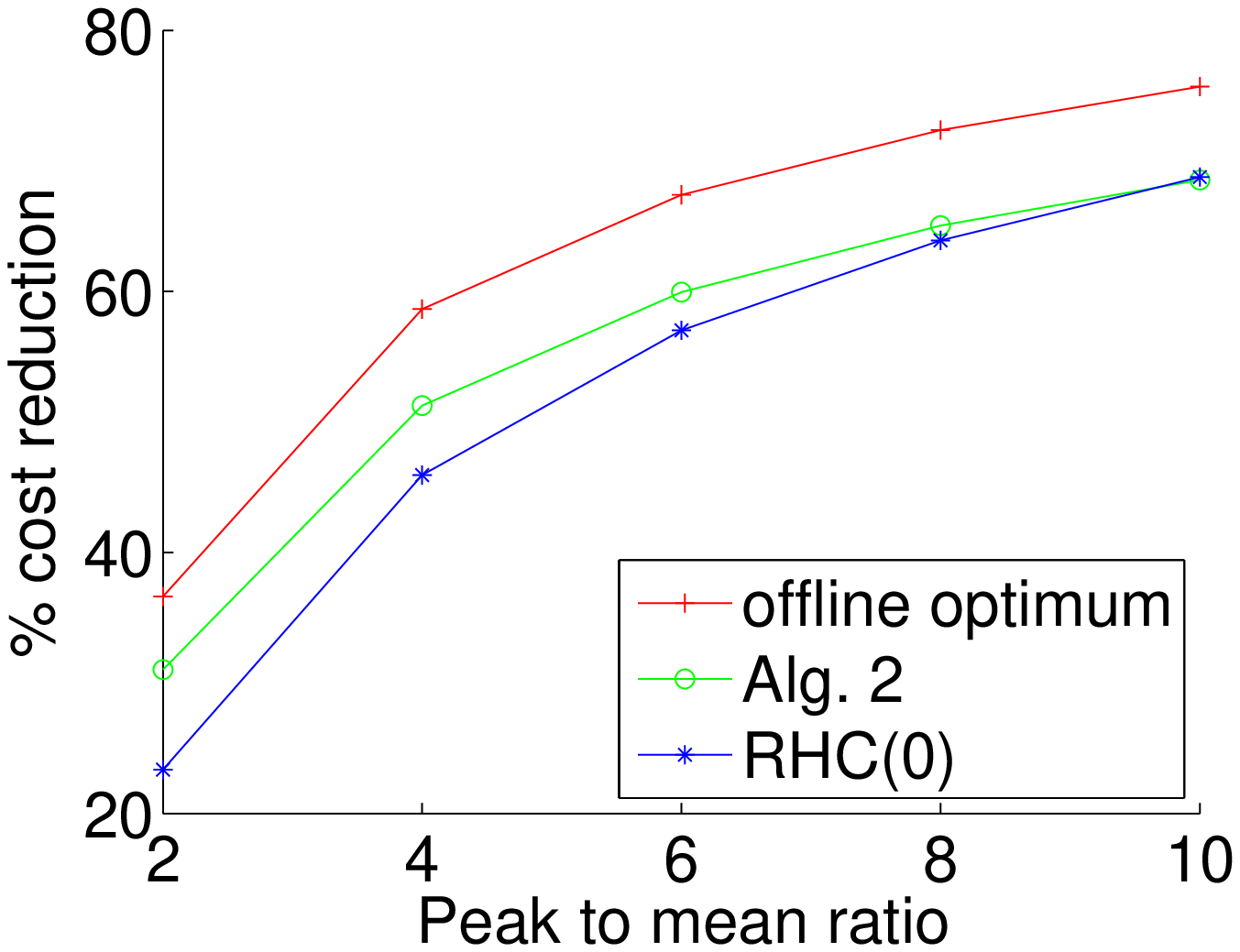}
  \caption{Impact of PMR for SSC}
  \label{fig:pmr}
\end{minipage}
\hfill
\begin{minipage}[t]{0.22\linewidth}
  \includegraphics[width=\textwidth]{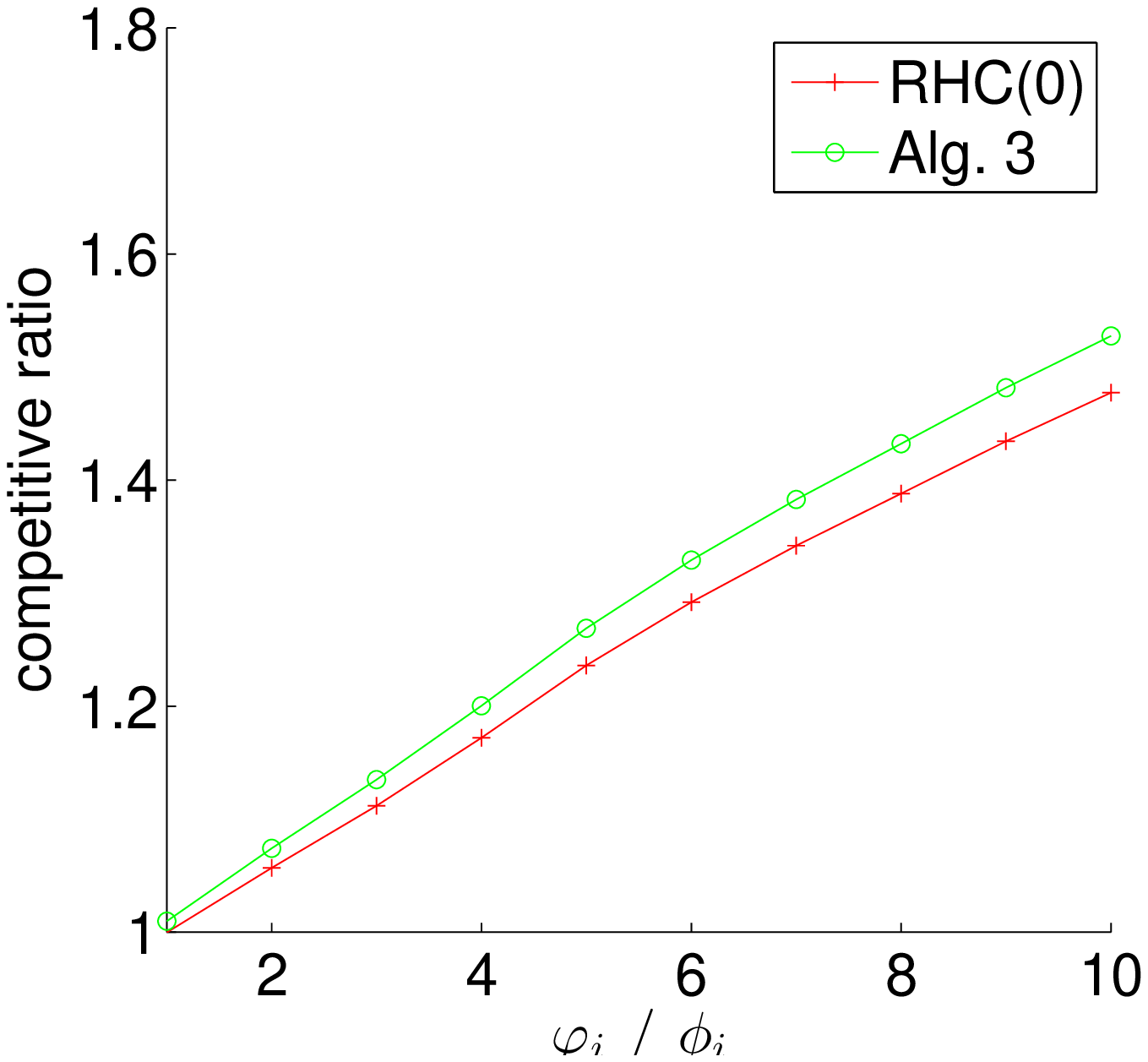}
  \caption{Impact of $\varphi_i/\phi_i$ on competitiveness for MSC}
  \label{fig:deployment2operational_msc}
\end{minipage}
\end{center}

\vspace{-7mm}
\end{figure*}

\section{Performance Evaluation}\label{sec:experiment}

We evaluate the performance of our online algorithms using trace-driven simulations. We compare their performance with the offline optimum, as well the RHC(0) algorithm in \cite{lin2012online},
 which exactly solves the ILP (\ref{eq2:total_cost}) in each time slot. 

\subsection{Settings}

\emph{Workload traces}: Like in \cite{lu2013simple} and \cite{lin2013dynamic}, we use a week of I/O traces from 1 RAID volume, CAMRESHMSA01-lvm0, at MSR Cambridge \cite{narayanan2008write}, representing activities in a service used by hundreds of users. We use the trace data as time-varying input traffic rates to the service chains. The trace data is normalized such that the peak load is 400Gbps, and the peak-to-mean ratio (PMR) is 4.27.

\emph{Cost benchmark}: We use the cost incurred in static provisioning of VNF instances in a datacenter as a benchmark, for computing cost savings of different algorithms in comparison. In the static provisioning, we assume that the datacenter has the complete workload information ahead of time and provisions a constant number of VNF instances over time according to the peak load.

\emph{Servers and VNFs}: 
For simplicity, we assume CPU capacity is the only bottleneck in our experiments. 
 There are 1000 servers and each of them has 16 physical CPU cores. There are four types of VNFs to compose service chains, namely firewall, load balancer, NAT and IDS. Following configurations in \cite{2015arXiv150306377F}, a firewall instance requires 4 CPU cores and can handle an incoming traffic rate of 900Mbps. A load balancer or NAT instance requires 2 CPU cores and can handle 900Mbps incoming traffic rate. An IDS instance requires 8 CPU cores and can handle 600Mbps incoming traffic rate.

\emph{Service chains}: We consider the following service chains: firewall$\rightarrow$NAT, firewall$\rightarrow$IDS and firewall$\rightarrow$IDS$\rightarrow$load balancer. We assume that a firewall filters 10\% of the input traffic and an IDS filters 20\% of the input traffic on average. Load balancers and NATs in the service chains do not change the flow rates.

\emph{VNF operational cost}: 
 We set $\phi_i$ proportionally to $C_i$, 
  and $\varphi_i/\phi_i \in [1,10]$. It is generally reasonable since deployment cost is on the order of the cost to run a VNF instance for several minutes \cite{lin2013dynamic}.

\subsection{Performance of Algorithms for a Single Service Chain}
The service chain is firewall$\rightarrow$IDS$\rightarrow$load balancer. 
 According to Alg.~\ref{alg:MITR} and our setup, with 1000 servers, the system can handle at most 886Gbps input traffic and in total 985 firewall instances, 1329 IDS instances and 709 load balancer instances are required to serve this traffic. One feasible MAX placement solution is to have 664 servers with 2 IDS instances each, 246 servers with 4 firewall instances  each, 88 servers with 8 load balancer instances  each, 1 server with 1 IDS instance, 1 firewall instances and 2 load balancer instances and 1 server with 3 load balancer instances.

\subsubsection{Impact of Unit Deployment Cost to Operational Cost}
 Under different values of $\varphi_i/\phi_i$, Fig. \ref{fig:deployment2operational} shows the cost saving by comparing three algorithms to the benchmark, our algorithm in Alg.~\ref{alg:online randomized}, offline optimum derived by solving problem (\ref{eq2:total_cost}) exactly using MOSEK \cite{mosek2010mosek}, and the RHC(0) algorithm, respectively. We can see that 
our proposed randomized online algorithm can save up to 70\% of the cost and it performs better than RHC(0) when $\varphi_i/\phi_i$ is greater than 4. Although RHC(0) and Alg.~\ref{alg:online randomized} have similar performance when $\varphi_i/\phi_i$ is less than 4, it is worth noting that Alg.~\ref{alg:online randomized} guarantees better performance in the worst case and is computationally tractable (RHC(0) involves solving ILPs exactly).  Also, the cost reduction is decreasing as $\varphi_i/\phi_i$ increases, {\em i.e.}, the smaller the deployment cost is, the more cost saving we can obtain. This is because a static provisioning approach will only pay a deployment cost at the very beginning but a right-sizing approach will pay an increasing deployment costs each time it decides to vary its deployment. Therefore, we could save more cost as technology continues to advance, resulting in the reduction of deployment cost.



We also investigate the impact of $\varphi_i/\phi_i$ on the competitive ratio of different algorithms. The results are shown in Fig. \ref{fig:d2o_comp}. We can see the ratio achieved by our online algorithm is quite stable as $\varphi_i/\phi_i$ changes, while RHC(0) performs worse as $\varphi_i/\phi_i$ increases. This also confirms a previous result from \cite{lin2012online}, which proves RHC(0) is ($1+\varphi_i/\phi_i$)-competitive.

\subsubsection{Impact of Peak-to-Mean Ratio (PMR)}
Intuitively, comparing to static provisioning, dynamic provisioning can achieve more cost saving when the input workload varies more significantly. Our following evaluation confirms this intuition. 
Similar to \cite{lin2013dynamic}, we generate the workload based on the MSR traces by scaling $\alpha(t)$ as $\overline{\alpha(t)} = K\alpha(t)^\gamma$, and adjust $\gamma$ and $K$ to keep the mean constant. We let PMR of the input traffic rates at different times vary from 2 to 10. Notice when PMR = 10, the maximal input traffic rate is 879Gbps, which is quite close to 886Gps. Fig. \ref{fig:pmr} shows that the cost saving increases from about 30\% at PMR=2 (which is common for workload in large datacenters) to about 67\% for higher PMRs (that are common in small to medium sized datacenters \cite{lu2013simple}).


\subsection{Performance of Algorithms for Multiple Service Chains}
We experiment with three service chains, namely firewall$\rightarrow$NAT, firewall$\rightarrow$IDS and firewall$\rightarrow$IDS$\rightarrow$load balancer. 
We compare the competitive ratios achieved by our proposed online algorithm in Alg.~\ref{alg:online msc} and RHC(0), computed against the offline optimum obtained by solving the offline ILP in (\ref{eq2:total_cost}) exactly.

Fig. \ref{fig:deployment2operational_msc} shows that the competitive ratios of Alg.~\ref{alg:online msc} are slightly larger than those of RHC(0). Considering the fact that Alg.~\ref{alg:online msc} does not require solving an ILP in each time slot while RHC(0) does, such a small loss of competitiveness is acceptable. We also observe that the competitive ratios are increasing as $\varphi_i/\phi_i$ becomes larger, which is consistent with our theoretical competitive ratio analysis.

%
%

\section{Concluding Remarks}\label{sec:conclude}
Network function virtualization provides a flexible way to deploy, operate and orchestrate network services with much less capital and operational expenses. Software middleboxes (e.g., ClickOS) are rapidly catching up with hardware middleboxes in performance. Network operators are already opting for NFV based solutions. We believe that our proposals for dynamic VNF provisioning and placement will have a significant impact on VNF management in the near future. Our model can be used to determine the optimal numbers of VNF instances and their optimal placement on servers, to optimize operational cost and resource utilization over the long run. Two solutions are designed: for a single service chain, we obtain a randomized online algorithm with competitive ratio $e/(e-1)$; for multiple service chains, we design a $(1+\max\limits_{i\in[I]}\frac{\varphi_i}{\phi_i})$-competitive online algorithm, relevant to the ratio of deployment cost over operational cost. Our trace driven simulations demonstrate that the overall cost can be reduced significantly by our dynamic VNF provisioning algorithms.


\bibliographystyle{IEEEtran}
\bibliography{main}

\begin{appendices}

\section{Proof of Theorem \ref{thm:lower bounds of VNFs}} \label{app:thm:lower_bounds_of_VNFs}
The total incoming traffic rate to all instances of VNF $i$ in service chain $s$ is $\bar{\lambda}_i^s\alpha^s(t)$, according to the generalized flow conservation constraint (\ref{eq:flow conservation}). Since one instance of VNF $i$  can maximally handle an input traffic rate of $b_i$, at least $\lceil \frac{\sum_{s\in[S]}\bar{\lambda}_i^s\alpha^s(t)}{b_i} \rceil$ instances are needed.

\section{Proof of Theorem \ref{thm:MITR}} \label{app:thm:MITR}
Because the server placement of an instance is decided by the removed element from the multiset and the element inserted into the multiset is decided by server placement of the removed instance, the set of server placement of VNF $i$ instances is always a subset of the initial $S_i$, which represents the positions of VNF instances to serve the maximum input traffic rate $\alpha^{max}$. Therefore, server capacity constraint (\ref{eq2:server CPU}) is always respected. Also, in a time slot, we either remove some elements from $S_i$ or insert some elements into $S_i$. Therefore, no migration of VNF instances occurs.

\section{Proof of Theorem \ref{thm:equal_offline_problems}} \label{app:thm:equal_offline_problems}
Since constraint (\ref{eq2:input}) is derived from constraint (\ref{eq:input}), (\ref{eq:flow conservation}) and (\ref{eq:VNF processing capacity}), any feasible solution to problem (\ref{eq1:total_cost}) is also a feasible solution to problem (\ref{eq2:total_cost}). On the other hand, given any feasible solution to problem (\ref{eq2:total_cost}), we can apply the proportional routing solution, which can guarantee all the flows are served, flow conservation constraint is respected and no VNF instance is over-committed, and obtain a feasible solution to problem (\ref{eq1:total_cost}). Since the two problems have the same objective function, we arrive at the conclusion that the two problems are equivalent.

\section{Proof of Theorem \ref{thm:polynomial}} \label{app:thm:polynomial}
Before we prove the polynomiality of Alg. 1, we will introduce the following lemma first, which is proved by \cite{goemans2014polynomiality}.

\label{sec:thm:maintheorem}
\begin{lemma} \label{thm:MainGeneralTheorem} \cite{goemans2014polynomiality}
Given polytopes $P, Q \subseteq R^L$, one can find a $y \in  \textrm{int.cone}(P \cap Z^L) \cap Q$ and
a vector $\lambda \in Z_{\geq0}^{P \cap Z^L}$ such that $y = \sum_{x \in P \cap Z^L} \lambda_x x$ in time
$\textrm{enc}(P)^{2^{O(L)}} \cdot \textrm{enc}(Q)^{O(1)}$, or decide that no such $y$ exists.
Moreover, the support of $\lambda$ is always bounded by $2^{2L+1}$.
\end{lemma}

In fact, by choosing $P = \{ {x \choose 1} \in R^{L+1}_{\geq 0} \mid \sum\limits_ic_{ir}x_i \leq C_r, \forall r \in [1,R] \}$, which is the set of feasible patterns, and $Q = \{ a\} \times [0,U]$, which is target state, we can decide in polynomial time, whether we can pack all the items in $U$ bins. Theorem~\ref{thm:polynomial} then follows using binary search.

\section{Proof of Theorem \ref{thm:online randomized}} \label{app:thm:online randomized}
Since the VNF placement scheme won't cause VNF instance migration and can guarantee constraint (\ref{eq2:server CPU}) is respected, problem (\ref{eq3:total_cost}) is equivalent to problem (\ref{eq2:total_cost}). Since the modified version of online randomized ski-rental algorithm can solve problem (\ref{eq3:total_cost}) with a competitive ratio of $e/(e-1)$ \cite{karlin1994competitive} and Alg. 2 combines the online ski-rental algorithm to VNF placement scheme, we arrive at the conclusion that Alg. 2 has a competitive ratio of $e/(e-1)$.

\section{Proof of Theorem \ref{thm2:MSC competitive ratio}} \label{app:thm2:MSC competitive ratio}
We have $[x_{ui}(t)-x_{ui}(t-1)] \leq x_{ui}(t)$. Therefore, $\phi_ix_{ui}(t) + \varphi_i[x_{ui}(t)-x_{ui}({t-1})]^+ \leq (1+\max\limits_{i\in[I]}\frac{\varphi_i}{\phi_i})\phi_ix_{ui}(t)$. Therefore, any algorithm which can guarantee to minimize $\sum\limits_{u\in [U]}\sum\limits_{i\in[I]}\phi_ix_{ui}(t)$ in each time slot has a competitive ratio of $(1+\max\limits_{i\in[I]}\frac{\varphi_i}{\phi_i})$.

\end{appendices}


%



\end{document}